\documentclass[11pt]{article}

\usepackage{cite} 
\usepackage{url}  
\usepackage{ifthen}
\usepackage{multicol}
\usepackage[utf8]{inputenc}
\usepackage{algorithm,algorithmic,citesort}
\usepackage{amsmath}
\usepackage{graphicx}
\usepackage{epsfig}
\usepackage{anysize}
\usepackage{setspace}

\newboolean{publ}

\newcommand{\utwi}[1]{\mbox{\boldmath $ #1$}}

\newcommand{\bA}{{\utwi{A}}}

\newcommand{\bI}{{\utwi{I}}}

\newcommand{\bM}{{\utwi{M}}}
\newcommand{\bN}{{\utwi{N}}}

\newcommand{\bP}{{\utwi{P}}}

\newcommand{\bR}{{\utwi{R}}}
\newcommand{\bS}{{\utwi{S}}}
\newcommand{\bT}{{\utwi{T}}}

\begin{document}
\title{Optimal enumeration of state space of finitely buffered
stochastic molecular networks and exact computation of steady state
landscape probability}

\author{Youfang Cao$^1$ \and Jie Liang$^{1,2}$ \\
    $^1$Shanghai Center for Systems Biomedicine (SCSB), \\
    Shanghai Jiao Tong University, Shanghai 200240, China.\\
    $^2$Department of Bioengineering, University of Illinois at
    Chicago, \\
    Chicago, IL 60612, USA
}

\date{Published on BMC Systems Biology on Mar. 29 2008 vol. 2, pp. 30}

\maketitle

\begin{abstract}

Stochasticity plays important roles in many molecular networks when
molecular concentrations are in the range of $0.1 \mu$M to $10 n$M
(about 100 to 10 copies in a cell).  The chemical master equation
provides a fundamental framework for studying these networks, and the
time-varying landscape probability distribution over the full
microstates, {\it i.e.}, the combination of copy numbers of molecular
species, provide a full characterization of the network dynamics.
A complete characterization of the space of the microstates is a
prerequisite for obtaining the full landscape probability distribution
of a network.  However, there are neither closed-form solutions nor
algorithms fully describing all microstates for a given molecular
network.

We have developed an algorithm that can exhaustively enumerate the
microstates of a molecular network of small copy numbers under the
condition that the net gain in newly synthesized molecules is smaller
than a predefined limit.  We also describe a simple method for
computing the exact mean or steady state landscape probability distribution
over microstates.  We show how the full landscape probability for the
gene networks of the self-regulating gene and the toggle-switch in the
steady state can be fully characterized.  We also give an example
using the MAPK cascade network.  Data and server will be available at
URL: {\tt http://scsb.sjtu.edu.cn/statespace}.

Our algorithm works for networks of small copy numbers buffered with a
finite copy number of net molecules that can be synthesized,
regardless of the reaction stoichiometry, and is optimal in both
storage and time complexity.  The algorithm can also be used to
calculate the rates of all transitions between microstates from given
reactions and reaction rates.  The buffer size is limited by the
available memory or disk storage.  Our algorithm is applicable to a
class of biological networks when the copy numbers of molecules are
small and the network is closed, or the network is open but the net
gain in newly synthesized molecules does not exceed a predefined
buffer capacity.  For these networks, our method allows full
stochastic characterization of the mean landscape probability
distribution, and the steady state when it exists.

\end{abstract}

\ifthenelse{\boolean{publ}}{\begin{multicols}{2}}{}

\section*{Background}

Networks of interacting biomolecules are at the heart of the
regulation of cellular processes, and stochasticity plays important
roles in many networks, including those responsible for gene
regulation, protein synthesis, and signal transduction
\cite{McAdams_PNASUSA97,Arkin_Genetics98,Hasty_PNASUSA00,Ozbudak_NG02,Levin_FL03}.
The stochasticity originates intrinsically from the small copy numbers
of the molecular species in a cell , which frequently occur when
molecular concentrations are in the range of $0.1 \mu$M to $1 n$M
(typically from about 100 to 10 copies in a cell)
\cite{Arkin_Genetics98,Morishita_JTB04}.  For example, the regulation
of transcriptions depends on the binding of often a few proteins to a
promoter site; the synthesis of protein peptides on ribosome involves
a small copy number of molecules; and patterns of cell differentiation
depend on initial small copy number events.  In these biological
processes, fluctuations due to the stochastic behavior intrinsic in
low copy number events play important roles.

The importance of stochasticity in cellular functions is well
recognized.  Studies of network models show that stochasticity is
important for magnifying signal, sharpening discrimination, and
inducing multistability
\cite{Paulsson_PRL00,Kepler_BJ01,Ozbudak_NG02,Zhou_PRL05,Samoilov_PNASUSA05,Volfson_Nature06,Mettetal_PNASUSA06,Morishita_BJ06,Mettetal_PNASUSA06}.
Understanding the stochastic nature and its consequences for cellular
processes involving molecular species of small copy numbers in a
network is an important problem.

A fundamental framework for studying the full stochasticity is the
chemical master equation \cite{Gillespie_JPC77,vanKampenBook}. Under
this framework, the combination of copy numbers of molecular species
defines the microscopic state of the molecular interactions in the
network.  By treating microscopic states of reactants explicitly,
linear and nonlinear reactions (such as synthesis, degradation,
dimeric binding, and multimerization) can all be effectively modeled
as transitions between microstates, with transition rates determined
by the physical properties of the molecules and the cell environment.
The probability distribution or potential landscape
\cite{AoKwonQian07_Complexity,Kim_PCB07,Schultz_JCP07} over %Qian
these microstates and its time-evolving behavior provide a full
description of the properties of a stochastic molecular network.

However, it is challenging to study a realistic system that involves a
nontrivial number of species of small copy numbers.  Analytical
solutions of the chemical master equation exist only for very simple
cases, such as self-regulating genes \cite{Hornos_PRESNSMP05}, and the
toggle-switch network under certain restrictions
\cite{Kepler_BJ01,Schultz_JCP07}.  Instead of solving the master
equation, a widely used method is to carry out Monte Carlo simulations
using the Gillespie algorithm \cite{Gillespie_JPC77}.  This method
generates samples from multiple runs of simulation, and statistics
properties are calculated from the simulation trajectories, which
provide characterizations of the network
\cite{Gillespie_JPC77,Morishita_BJ06,Gillespie_JCP2003,Cao_JCP2004}.
This approach has found wide applications, although it cannot
guarantee a full account of stochasticity, as this method usually does
not generate an exhaustive number of trajectories that cover all
possible locations in the probability landscape.  In addition, as
Monte Carlo simulations follow high probability paths, it is
especially challenging to sample adequately rare and critical events
that may be important in determining cellular fate.  It is also
difficult to determine whether a simulation is extensive enough to
obtain accurate statistics.  The amount of computation necessary to
obtain an accurate result may be too large to be completed in a
reasonable amount of time, especially when the time scales of the
various react\ ions involved are very different \cite{Kepler_BJ01}.
To address these issues, Gillespie, Petzold, and colleagues further
developed numerical methods for speeding up the stochastic simulation
\cite{Gillespie_JCP2003,Cao_JCP2004}.  Munsky and Khammash developed a
method to approximate the solution of chemical master equation by
projecting the whole state space of the system to a finite space
\cite{Munsky_JCP2006}.  Samant and Vlachos developed a multiscale
Monte Carlo method for stiff systems where partial equilibrium occurs
\cite{Samant_JCP05}.  An alternative approach is to approximate the
master equation using, for example, Fokker-Planck or Langevin
equations \cite{vanKampenBook}. These are obtained by adding
stochastic terms (often Gaussian) to a deterministic equation
\cite{Schultz_JCP07,Zhu_JBCB04,Mettetal_PNASUSA06}. Salis and
Kaznessis improved the stochastic simulation method by partitioning
the system into components with fast and slow reactions. The fast
reactions are approximated by the Langevin equations, and the slow
reactions are analyzed by stochastic Monte Carlo simulations
\cite{Salis_JCP2005}.

A complete identification and characterization of the space of the
microstates is a prerequisite for obtaining the full landscape
probability distribution of a network.  However, the state space of a
network currently cannot be fully characterized in general.  There is
neither closed-form solution, nor computational algorithm describing
the full state space.  In this paper, we study the problem of
enumerating the state space of a molecular network with small copy
numbers of molecular species.

A naive method is to predefine the maximum copy number of the
reactants, and bound the state space by the product of the maximum
numbers.  However, the size of state spaces estimated by this naive
approach will be inflated to enormity. For example, if there are 16
species, and there is a total a maximum of 33 molecules in the whole
system, this naive method does not take into consideration of the
details of the network, and the state space will be estimated to be in
the order of $(33+1)^{16} = 3.19\times10^{24}$
states.  This naive method is intrinsically inefficient: There may be
many states which may never occur. For some states, no reactions
may occur and therefore are not needed.  For others, no reactions can
lead to them under the specified initial condition.  An alternative
approach is carrying out simulation.  One can simply follow explicitly
simulated reaction events to whatever microstates of copy numbers the
system reaches.  However, this approach cannot guarantee that all
reachable states will be explored, therefore cannot guarantee full
characterization of rare events.

In this study, we develop an optimal algorithm that gives full
description of the state space and the set of transitions.  Our method
works for networks of small copy numbers under the condition that the
net gain in newly synthesized molecules in the network does not exceed
a predefined finite number. Our algorithm is optimal in both memory
requirement and in time complexity. All states reachable from a given
initial condition will be accounted for by our method, and no
irrelevant states will be included. All possible transitions will be
recorded, and no infeasible transitions will be ever attempted.  As a
result, our algorithm can generate the full state-transition matrix
under the framework of the chemical master equation. This matrix is
compact without any redundant information. It is also of the minimal
size.  In addition, the computational time is optimal, up to a
constant.  We also describe how to obtain the mean landscape
probability distribution over the enumerated state space of a network,
which is the same as the landscape distribution of the steady state
when it exists.

This paper is organized as follows. We first describe how our method
can be applied to the simple examples of a self-regulating gene, a
toggle-switch network, and the more complex example of the MAPK
network. This is followed by conclusion and discussion. We finally
describe the technical details of the algorithm for enumerating the
space of microstates, and introduce a simple method for
computing the steady state landscape probability distribution.

\section*{Results and Discussion}
\subsection*{Molecular network models}
We apply our algorithm to three network models: the self-regulating
 gene, the small toggle-switch network, and the MAPK cascade network.

\subsubsection*{Self-regulating gene}
Regulating the expression of even a single gene is a complex process.
We study the network of an idealized self-regulating gene
(Fig~\ref{fig:selfreg-top}a and b).  As a basic unit in biological
genetic networks, it consists of only one gene, and is the simplest
molecular network.  We follow the study of Schultz {\it et al} and
assume that the dominant form of regulation is the binding and
unbinding of transcription factors to the operator site, which changes
the rate of transcription initiation \cite{Schultz_JCP07}.  In this
model, there are several stochastic processes: the synthesis and
degradation of the protein transcription factor at the reaction rate
constants of $s_0$ (or $s_1$) and $d$, respectively, and the binding and
unbinding of the operator site of DNA by the transcription factor at
the reaction rate constants of $b$ and $u$, respectively.  These
processes are illustrated in Fig~\ref{fig:selfreg-top}b.  The binding
state of the operator site is either ``on/unbound'' (state 1), or
``off/bound'' (state 0). The synthesis rate of transcription factor is
either $s_0$ or $s_1$, depending on the binding state of the operator
site.

We first calculate the state spaces.  We use the same initial
condition of 1 copy of unbound gene, 0 copies of transcription factor
and bound gene, and set the buffer size to allow different copy
numbers of protein transcription factor to be synthesized.  As there
is only one copy of the gene in this model \cite{Schultz_JCP07}, the
size of the state space increases with the copy number of the protein
transcription factor that can be synthesized.  Our results show that
when the buffer capacity takes the value of 100, 1,000, and 10,000,
the size of the state space is 201, 2,001, and 20,001, respectively.
In this model, the size of the state space scales linearly with the
copy number of the protein synthesized.  In biological condition, the
copy number of a transcription factor rarely exceeds 100.

We then calculate the exact steady state probability distribution
over the microstates of the self-regulating gene, namely, the
exact steady state density function of different states of copy
numbers of the transcription factor.  In our calculation, the
parameter values are chosen as $ u=d/10$ and $b=d/250$, in units of
degradation rate $d$, following reference \cite{Schultz_JCP07}.
 The steady state distributions $\bP$ at
different values of synthesis rates in on/unbound and off/bound states
$s_1$ and $s_0$ are computed exactly and are shown in
Fig~\ref{fig:selfreg-prob} for the case of buffer size of 1,010 for
illustration.  Here the marginal probability of having a specific
number of free proteins in the system is plotted, regardless whether
the gene is in off/bound or in on/unbound state.  Following reference
\cite{Schultz_JCP07}, we use three different network conditions:
$(s_0, s_1) = (50, 10),\, (50, 50)$, and $(10, 50)$ in units of
degradation rate $d$, respectively.  When the on/unbound state
synthesis rate $s_1$ is greater, the network is self-repressing.  When
the off/bound synthesis rate $s_0$ is greater, the network is
self-activating.

Our results and the results of Schultz {\it et al}\/ obtained from
multiple runs of Gillespie simulations are identical
\cite{Schultz_JCP07}.  As pointed out in \cite{Schultz_JCP07}, the
self-repressing and the self-activating genes can have overall similar
distributions.  This can be explained by the fact that the combined
synthesis rate of the protein $s_0 + s_1 = 60$ is the same in both
cases (front profile and back profile in Fig~\ref{fig:selfreg-prob}).
Closer examination shows that in the case of the self-repressing gene
network ($s_0 = 10$ and $s_1 = 50$, front profile), the first peak of
probability at smaller copy number of the free protein is lower, and
the second peak at higher copy number is larger when compared to the
distribution of the self-activating gene ($s_0 = 50$, and $s_1 = 10$,
Fig~\ref{fig:selfreg-prob}, back profile).  That is, the
self-repressing network has a higher probability in producing more
free proteins than the self-activating network. This can be explained
by the difference between the protein-DNA binding rate $b$ and
unbinding rate $u $. In this model network, unbinding rate $u = d/10$
is 25 times greater than the binding rate $b = d/250$.  As a result,
this gene is more likely to stay in the unbound state.  Since the
self-repressing network has a higher synthesis rate in unbound state
($s_1 = 50 > s_0 = 10$), it will produce more free proteins.  This
results in an overall slightly higher probability for larger number of
free proteins for self-repressing network.  This small difference in
probability distribution is also observed in \cite{Schultz_JCP07}.  As
pointed out previously in \cite{Schultz_JCP07}, when both synthesis
rates are equal ($s_0 = s_1 = 50$), the binding state transition do
not change the synthesis/degradation process, and the network is a
simple birth/death process, with a Gaussian probability distribution
for protein number centered at $s_0=s_1$ (Fig~\ref{fig:selfreg-prob},
middle profile).

\subsubsection*{Toggle switch}
A toggle switch is a small network consisting of two genes, A and B.
The protein product of each represses the other gene. Toggle switch is
the smallest genetic network that can present bistability.  The
insightful study of Schultz {\it et al\/} provided detailed analysis
of the stochastic behavior of this model network \cite{Schultz_JCP07}.
To facilitate direct comparison, we adopt the same toggle-switch
model developed by these authors (Eqns 5--8 in reference
\cite{Schultz_JCP07}).
The molecular species and the network topology
are shown in Fig~\ref{fig:toggle-top}a.  There are a number of
stochastic processes: the synthesis and degradation of proteins A and
B, with reaction constants denoted as $s$ and $d$, respectively;
the binding and unbinding of the operator site of one gene by the
protein products of the other gene at rate $b$ and $u$, respectively
(Fig~\ref{fig:toggle-top}b).  The binding states of the two operator
sites are ``on-on/unbound-unbound'' (state 11 for gene A and gene B),
``on-off/unbound-bound'' (state 10), ``off-on/bound-unbound'' (state
01), and ``off-off/bound-bound'' (state 00).  The synthesis rates of
both proteins A and B depend on the binding state of the operator
sites.  The toggle switch model used in this study and all possible
chemical reactions in the model are extracted directly from the master
equations in \cite{Schultz_JCP07}. In this model, no dimerizations are
explicitly modeled, and the model assumes that binding of two proteins
to the operator site simultaneously.  This is a valid approximation
when the dimerization reaction is fast compared to all other reactions
\cite{Kepler_BJ01}.  Even for this simple network, except for the
special cases when ``fast transition'' between on- and off- operator
states and ``small noise'' of high molecular concentration conditions
are assumed, no exact solutions are known
\cite{Kepler_BJ01,Schultz_JCP07}.

We first calculate the state spaces under the initial condition of 1
copy of unbound gene A, 1 copy of unbound gene B, 0 copies of bound
gene A and bound gene B, and 0 copies of their protein products.  We
set the buffer size to different copies of total protein A and protein
B combined that can be synthesized.  When the buffer capacity is 20,
the size of the state space is 764.  At buffer capacity of 200, 400,
and 800 copies of proteins, the size increases to 79,604, 319,204, and
1,278,404, respectively.

We then calculate the exact steady state landscape probability of the
toggle-switch network, namely, the exact steady state density function
of different microstates of copy numbers of products of gene A and
gene B.  The steady state distributions
$\bP$ are shown in Fig~\ref{fig:toggle-prob} for the case of buffer
size of 300.  In our calculation, the parameter values are chosen as
$s=100d, u=d/10$, and $b=d/100,000$, in units of degradation rate $d$.
These are the same as those used in reference \cite{Schultz_JCP07}.

It is clear that a toggle switch has four different states,
corresponding to the ``on/on'', ``on/off'', ``off/on'' and ``off/off''
states.  At the chosen parameter condition, the toggle/switch exhibits
clear bi-stability, namely, it has high probabilities for the
``on/off'' and ``off/on'' states, but has a low probability for the
``on/on'' state.  The ``off/off'' state is severely suppressed.  Our
results are identical with the results of Schultz {\it et al}\/
obtained from multiple runs of Gillespie simulations
\cite{Schultz_JCP07}.

\subsubsection*{MAPK network}
MAPK cascade network plays important role in signal transduction.
Here our purpose is to explore how to apply our algorithm to more
realistic network model.  Our goal in this paper is not to study the
the stochastic nature and the dynamic behavior of MAPK network.

The MAPK cascade network (BioModels ID: BIOMD0000000028) is taken from the BioModels database at EBI
\cite{Markevich_JCB04,BIOMODEL}. The molecular species and reactions are
extracted from the SBML (Systems Biology Markup Language) model file.
This network contains 16 molecular species with 17 reactions
\cite{Markevich_JCB04}.  As there is no synthesis reaction, this
particular network model is a closed system.  Abbreviations used in this
model are listed in Table~\ref{tab:abbrev}.  Fig~\ref{fig:model} shows
the topology of the model. All 16 molecular species are labeled with
numbers from 1 to 16. Among them, MEK (triangles in
Fig~\ref{fig:model}) and MKP3 (squares) are the key enzymes catalyzing
all phosphorylation and dephosphorylation reactions in this
network. The rest of the molecular species are substrates,
intermediates, and products of MEK and MKP3 induced reactions. Most of
the reactions in this model (14 of 17) are second-order.

\noindent {\bf Simple initial conditions.}
We generate the state spaces of the MAPK network for different
initial conditions and record their sizes.
We first increase the copy number for one species from 1 to
20, and record the size of resulting state space, while keeping the
copy numbers of all other species to $0$.  We repeat this process
for each of the 16 molecular species in turn.  Altogether, we have $16
\times 20 = 320$ data points of sizes of the state space
(Fig~\ref{fig:space}).

It is clear that different molecular species in this model affect the
size of the state space differently.  Increasing the copy number of
M-MEK-Y, M-MEK-T, and Mpp-MKP3 molecules (species 9, 10 and 11, in
bold fonts in Table 1) lead to large state spaces (size $888,030$ at
20 copies, Fig~\ref{fig:space}), while the initial conditions of 20
copies of any other species result in modest state spaces.  For
example, species 7, 8, 15 and 16 when given 20 copies have a
state-space size of 231.  For species 1--6 (M, MpY, MpT, Mpp, MEK,
MKP3), no reactions can occur at these initial conditions, and the
state space contains only the the initial state.

The state space for each of the 320 initial conditions can be computed
within one minute.  We further found that when any of $S_9, S_{10}$, or
$S_{11}$ has an initial copy of 28 and all others 0 copies,
the state spaces increases to 6,724,520, and the computing
time also increase, although all can be computed within 10 minutes on
a Linux workstation.

\noindent {\bf Biological initial conditions.}  We further calculate
sizes of the state spaces with several biologically plausible initial
conditions, in which species M, MEK and MKP3 are all given an equal
number of $i$ copies, while all the other species start with zero
copies.  We increase $i$ from $1$ to $11$.  These initial conditions
correspond to a total number ranging from $3\times 1 = 3$ copies to $3\times
11= 33$ copies of molecules of three species in the network.  The size
of the state space increases with the copy numbers.  When there are 1
copy of M, MEK, and MKP3 each, the size of the state space is 14.  For
5, 10, and 11 copies of M, MEK, and MKP3 each, the size increases to
8,568, 1,144,066, and 2,496,144, respectively.  The computation of the state
space at $i=10$ and $i=11$ requires 156 seconds and 589 seconds of CPU
time on a Linux desktop machine, respectively.

\noindent {\bf Steady state distribution.}
We compute the steady state probability distributions of the
microstates of the MAPK network at the initial condition of 10 copies
each of M, MEK and MKP3.  That is, we obtain the exact steady state
density function of different microstates of all possible 1,144,066
combinations of different copy numbers of the 16 molecular species in
the MAPK network. The computation is efficient. At this initial condition, the dimension of the Markovian
transition matrix $M$ is $1,144,066\times 1,144,066$, with $14,574,406$
number of non-zero elements.
It takes 1,341 seconds (about 23 minutes) of CPU time to compute
the steady state probability distribution on a Linux workstation.

As it is impossible to directly visualize the landscape density
distribution in a 16-dimensional space, for ease of visualization, we
plot the marginal distribution of different combinations of copy
numbers of extracellular signal-regulated kinase (ERK) in
unphosphorylated state, in single phosphorylated state, and in dual
phosphorylated state.  Specifically, we plot the marginal
probabilities of different copy numbers of unphosphorylated ERK (M),
and ERK with either Y or T site phosphorylated (Mp, including both MpY
and MpT), after integrating different copy numbers of all other 14
molecular species in Fig~\ref{fig:mapk-steady}a.  We plot the marginal
distribution of different copy numbers of unphosphorylated ERK (M),
and ERK with both Y or T site phosphorylated (Mpp)
Fig~\ref{fig:mapk-steady}b.  We plot the marginal distribution of
different copy numbers of uni-phosphorylated ERK with either Y or T
site phosphorylated (Mp, including both MpY and MpT), and ERK with
both Y or T site phosphorylated (Mpp) in Fig~\ref{fig:mapk-steady}c.
At this parameter condition, the steady state distribution has a
single peak centered around two copies of unphosphorylated ERK (M),
two copies of uni-phosphorylated ERK (Mp), and zero copy of dual
phosphorylated ERK (Mpp).

\section*{Conclusion}
Stochasticity plays important roles in molecular networks for
processes involving small copy numbers of molecules.  Models
of molecular networks based on macroscopic reaction rates and coupled
ordinary differential equations are not applicable in these cases, as
they can only model high concentrations of interacting molecules with
negligible fluctuations.

The stochastic nature of molecular interactions at low copy numbers
can be fully characterized if the time-varying landscape probability
distribution on all of the microstates of a molecular network can be
computed.  This is a difficult task, as the state space of the
combination of the copy numbers of molecular species needs to be
explicitly enumerated, the probability distribution over these
microstates and changes of this distribution across many decades of
time scale need to be fully computed.

In this study, we have developed an algorithm to enumerate the state
space of a molecular network of small copy numbers with a buffer
containing a finite number of molecules that can be synthesized.  It
can also be used to find all possible transitions between states, and
to compute the transition rates between these states.  We also
demonstrate how to obtain the steady state probability distribution
based on the enumerated states when it exists.

Our example of the toggle-switch network shows that this method can be
used to study the rise of important network properties such as
bistability.  The enumeration of the full state space of the MAPK
cascade network at various initial conditions demonstrate that our method
can be used to study a realistic network of nontrivial size, which is
more complicated than the simple networks that are often studied for
full stochasticity.  Although naively the state space at the initial
condition of each of 11 copies of unphosphorylated,
uniphosphorylated, and biphosphorylated ERK kinase might be as high as
$(33+1)^{16} = 3.19\times10^{24}$, a truly astronomical size, our
method showed that the relevant space is only about $2.50\times 10^6$,
which is amenable for computation using a desktop computer.

Our method is applicable to study various carefully constructed model
network systems.  It complements the Monte Carlo simulation method, as
it can be used to characterize the full probability landscape of
networks with enumerable state space.  For example, it will allow the
calculation of the probabilities of the occurrence of rare and
critical events.  For theoretical studies, one can predefine a fixed
number of net molecules that can be synthesized, and investigate the
nature of the landscape probability distribution.  This is similar to
the studies of semi-grand canonical ensemble in statistical physics
\cite{Hill-87}. Exact characterization of probability landscape is
useful, as most network studies are based on stochastic simulation,
and relative little is known at the level of the full stochastic
landscape probability distribution, even for simple toy systems.  For
example, analytical solutions to the simple toggle switch model is
known only when the model parameters follows the restrictions of small
noise and fast transition \cite{Kepler_BJ01,Schultz_JCP07}.  We
believe our method can be used to study well designed model systems
beyond self-regulating genes and simple toggle switches, and the exact
results obtained will be helpful for understanding the basic
properties and design principles governing stochastic networks.  A
useful analogy to illustrate the utility of such model studies can be
found in the field of protein folding, where a large number of studies
using simple short chain HP lattice models revealed remarkable
insights about how complex proteins fold
\cite{Dill90,Socci94_JCP,Sali94_Nature,DillBrombergYue95_PS,Klimov96_PRL,Shakh98_FD,Ozkan1_NSB,Kachalo06-PRL}.

Our method can also be applied to more realistic biological networks,
such as the MAPK network model, which is a closed system according to
the annotated BioModels database \cite{BIOMODEL}.  Such closed systems could arise when
one focus on a submodule of a larger network. For the majority of
realistic networks which are open systems, an important determining
factor of the applicability of our method is the limit of the capacity
of a buffer, which has to be greater than the maximum copy number of
the net gain in protein molecules that can be synthesized.  In a cell,
this maximum number is determined by the life time of the cell, and
the net synthesis rate of protein molecules.  The latter depends on
both protein synthesis and degradation rates. A simple approach is to
estimate the net number of protein molecules that can be synthesized
during the life time of a cell. For example, the lifespan of an {\it
E.\ coli\/} cell is about 30 minutes \cite{Little_EJ99}.  Estimation
based on the rate limiting processes of transcription initialization
and elongation indicate that the protein synthesis rate ranges from
$0.0077/s$ (for the C1 protein) \cite{Li_PNASUSA97,Arkin_Genetics98}
to $0.0534/s$ for the Cro protein
\cite{Hawley_PNASUSA80,Hawley_JMB82,Arkin_Genetics98} in the lambda
phage system. Their degradation rates are about $0.0007/s$ and
$0.0025/s$, respectively \cite{Arkin_Genetics98}.  This suggests that
a useful bound of the copy number of newly synthesized molecules for
studying the lambda switch network system could be in the order of
150-200 copies under reasonable initial conditions. Naturally, the
exact number will depend on the details of the chosen network model
and the parameter values.  For example, models of cells under stress
with retarded synthetic rates may require a relatively small buffer
capacity.

In this study, we have described a method to compute the steady state
landscape probability distribution.  Steady state distribution is of
general interests when it exists, as has been shown in previous
studies \cite{Schultz_JCP07,Kim_PCB07}.  For realistic network,
another approach is to compute the time-dependent dynamic change of
landscape probability distribution, using techniques such as those
used in \cite{Kachalo06-PRL}.  We will describe this approach in more
details in future studies.

As the number of molecular species and their copy numbers increase,
the state space will eventually become prohibitively large for
explicit computation even with an optimal algorithm.  In these cases,
our method can be used to select important states and to control error
bounds at a specific tolerance for developing approximation methods,
an approach well demonstrated in \cite{Munsky_JCP2006}.

\section*{Methods}

\subsection*{The Algorithm}

Suppose we have a model of a biological network, which contains $m$
molecular species and can have $n$ reactions.  Given an initial
condition, namely, the copy numbers of each of the $m$ molecular
species, we aim to calculate all states that the biological system can
reach starting from this initial condition, under the condition that
the net number of molecules that can be synthesized does not exceed a
predefined limit.  These states collectively constitute the state
space of the network under this initial condition.

Formally, we have a model of a biological network $\bN =(\bM, \bR)$,
with $m+1$ number of molecular species: $\bM =(M_1, \ldots M_{m+1})$,
and $n$ reactions: $\bR=\{R_1, \ldots, R_n\}$.  Here $m$ of the
species are from the network.  A buffer of predefined capacity is used
to represent a pool of virtual molecules for open systems, from which
synthesis reactions can generate new molecules, and to which
degradation reactions can deposit molecules removed from the network.
We use the $m+1$-th species to represent this buffer pool.  The
combinations of copy numbers of all molecular species $\bS = (c_1,
\ldots, c_m, c_{m+1})$ form the microstate of the system, where
$c_{m+1}$ denotes the number of net new molecules that can still be
synthesized at this state.  A reaction can involve an arbitrary number
($\ge 1$ and $\le m$) of molecular species as reactants and/or
products, with any arbitrary positive integer coefficient ({\it i.e.},
arbitrary stoichiometry).  Synthesis reaction is allowed to occur only
if the buffer pool is not exhausted, namely, only if $c_{m+1}>0$.  The
set of all possible states $\bS$ that can be reached from an initial
condition following these rules constitute the state space of the
system: $\mathcal{X} = \{\bS\}$. The set of allowed transitions is
$\bT = \{t_{i\,j}\}$.  We are given with an initial condition:
$\bS^{t=0} = (c_1^{0}, c_2^{0}, \ldots, c_m^{0}, c_{m+1}^0)$, where
$c_i^{0}$ is the initial copy number of the $i$-th molecular species
at time $t=0$, and $c_{m+1}^0 = B$ is the predefined buffer size.  The
maximum copy number of net gain in newly synthesized molecules of the
system is restricted by this constant $B$.  Our aim is to enumerate
the state space $\mathcal{X}$ under this given initial condition.

The algorithm is written as Algorithm 1 (see Appendix).  It performs the following
computation: After initialization, we start with the initial state
$S^{t=0}$. We examine each reaction in turn to determine if this
reaction can occur for the current state.  If so, and if the buffer is
not exhausted, we generate the state that this reaction leads to.  If
the newly generated state was not encountered before, we add it to our
collection of states for the state space, and declare it as a new
state.  We repeat this for all new states, which is maintained by a
stack data structure.  This terminates when all new states are
exhausted.

In this algorithm, a stack data structure is used.  Description of the
stack data structure can be found in computer science textbooks such
as \cite{Cormen90}.  A stack is used here to store individual
states. These states are ``{\tt Push}''ed onto the stack: If we encounter a
previously unseen state, we create it and push it onto the stack so
further calculations on this state can be carried out at a later
stage.  We use the ``{\tt Pop}'' operation to obtain a state previously
stored on the stack to carry out these calculations. In this case, we
pop a state to examine what reactions can occur and what other states
these reactions can lead to.

We can compute the transition coefficient $\{a_{i,j}\}$ between two
microstates $\bS_i$ and $\bS_j$ using Algorithm 2  (see Appendix) following the
approach outlined in references
\cite{Gillespie_JPC77,Munsky_JCP2006,Schultz_JCP07}.
We give further details in later sections on how this is done for the
three networks studied here.

\subsubsection*{Correctness and optimality}
The state space and the transitions under a given initial condition
can be considered as a directed graph $G=(\mathcal{X}, \bT)$, in which
vertice are the state vectors, {\it i.e.}, the set of reachable states
$\mathcal{X}$, or the $m+1$-tuples of copy numbers of the $m+1$
molecular species, including the buffer. Edges are the set of allowed
transitions $\bT$ between the states, {\it i.e.}, reactions connecting
two state vertice. Two vertice $\bS_i \in \mathcal{X}$ and $\bS_j \in
\mathcal{X}$ are connected by a directed edge $t_{i,j} \in \bT$ if and
only if $\bS_i$ can be transformed to $\bS_j$ through a reaction.  Any
reachable state can be transformed from the initial state by one or
more steps of reactions, and the directed graph $G$ is a connected
graph.

Our algorithm implicitly generates this graph $G$.  Because the set of
reactions $\bR$ is finite, $G$ has a finite tree-width at any finite
steps away from the initial condition.  Assume the algorithm will not
terminate in finite steps.  Since in this algorithm each state is only
visited no more than twice, $G$ must have an unlimited depth. That is,
there must exist a path $p$ in the graph $G$ that starts from the
initial state and extends to infinite. Therefore  $p$ must contain an infinite
number of different states. This is impossible for any given initial
condition, as each molecular species
has a limited initial copy number, and the size of the buffer limits
the number of new molecules that can be synthesized in open systems.
The algorithm therefore must terminate.

This algorithm gives correct answers, assuming that the newly
synthesized molecules does not exceed the predefined buffer capacity.
This is because all states visited in the algorithm can be reached from
the initial condition, and all visited states is actually reached as
each is brought to by a chemical reaction.  In addition, all reachable
states will be visited, as the algorithm test at each state all
possible reactions, and will only terminates when all new states are
exhausted. It is easy to see all possible transitions between states
will be recorded.

The time complexity of our algorithm is optimal.  Since only unseen
state will be pushed onto the stack, every state is pushed and popped
at most once, and each state will be generated/visited at most twice
before it is popped from the stack. As access to each state and to
push/pop operations take $O(1)$ time, the total time required for the
stack operations is $O(|\mathcal{X}|)$.  As the algorithm examines each of the
$n$ reactions for each reached state, the complexity of total time
required is $O(n|\mathcal{X}|)$, where $n$ is usually a modest constant ({\it
e.g.} $<50$).  Based on the same argument, it is also easy to see that
the algorithm is optimal in storage, as only valid states and valid
transitions are recorded.

\subsubsection*{Computing mean and steady state probability distribution}
We can calculate the expected landscape probability distribution over
the microstates, namely, the exact mean density function of different
microstates of copy numbers in the network. It is the same as the
steady state probability distribution function if the steady state
exists.  Instead of calculating the time trajectories of changes in
the probability distribution and wait until it reaches equilibrium, we
use a simpler approach applicable to networks in which a steady state
exists.  Following Kachalo {\it et al} \cite{Kachalo06-PRL}, we obtain
the Markovian state transition matrix $\bM$ from the reaction rate
matrix $\bA$: $ \bM = \bI + \bA \cdot \Delta t$, where $\bI$ is the
identity matrix, and $\Delta t$ is the discrete time unit and is
chosen to be $1$.  The probability distribution function $\bP$ of the
microstates can be obtained by solving the equation $\bP = \bM \bP$.
The calculation of the steady state distribution $\bP$ is not
sensitive to the precise choice of the discrete time unit $\Delta t$.
The steady state distribution corresponds to the eigenvector of $\bM$
with eigenvalue of 1.  We use the Arnoldi method implemented in the
software {\sc Arpack} to compute the steady state distribution $\bP$
\cite{Lehoucq_Arpack}.

\subsection*{Computing transition coefficients}
The transition coefficient between different states connected by a
reaction is calculated by multiplying the intrinsic rate of this
reaction with the reaction order dependent combination number of
copies of reactants in the ``before'' state \cite{Gillespie_JPC77}.
We provide more details using examples from the three networks.

{\bf Self-regulating gene.}
Suppose the first order reaction
$$
Protein \xrightarrow{d}\emptyset
$$ enables the transition of the system from the microstate $i$ to
$j$. This reaction denotes the degradation of the protein molecule at
an intrinsic rate of $d$. The stoichiometry of this reaction dictates
that the copy number of protein $n_{p,\,j}$ in the ``after'' state $j$
is one less than the copy number $n_{p,\,i}$ in the ``before'' state $i$.
From the reaction formula, the transition coefficient $a_{i,j}$ for
the matrix $\bA$ is calculated as:
$$
a_{i,j} = d \cdot n_{p,\,i}.
$$ Recall that since a microstate is uniquely determined by the
combination of copy numbers of all molecular species, $n_{p,\,i}$
therefore is known as a state attribute.

For the second order reaction
$$
Protein + Gene \xrightarrow{b} BoundGene,
$$
the transition coefficient connecting the ``before''
state $i$ to the ``after'' state $j$ can be computed as:
$$
a_{i,j} = b \cdot n_{p,i} \cdot n_{g,i},
$$ where $b$ is the intrinsic reaction rate, $n_{p,\,i}$ is the protein
copy number at state $i$, and $ n_{g\,,i}$ is the copy number
of gene in state $i$, which is 1, as we assume there is only one copy
of the gene in this network model.

We can similarly compute the transition coefficient $a_{i,\,j}$ for the
reaction
$$
BoundGene \xrightarrow{u} Protein+Gene
$$
as $a_{i,\,j} = u
\cdot n_{bg,\,i}$, where $ n_{bg,\,i}$ is the number of bound gene in the ``before'' state,
which takes the value of 0 or 1 in this model, depending on whether
the gene is in protein-free or in protein-bound state.  For the simpler
reaction:
$$
BoundGene \xrightarrow{s_0} Protein,
$$
we have $a_{i,\,j}
= s_0 \cdot n_{bg,\,i}$, where $n_{bg,\,i}$ is the number of bound
gene in state $i$, which takes the value of 0 or 1. For the synthetic
reaction of
$$
Gene \xrightarrow{s_1} Protein,
$$
we have $a_{i\,j} = s_1 \cdot n_{g,\,i}$.

We have described how to compute the transition coefficient for all
reactions in the represser gene network.  In Algorithm 2, we can
compute the transition coefficient $a_{i,j}$ based on the formula of
the reaction leading from state $i$ to state $j$.

{\bf Toggle switch.}  For the third order reaction
$$
2\times ProteinA + GeneB \xrightarrow{b} BoundGeneB,
$$
the transition coefficient $a_{i,j}$ can be computed as
$$
a_{i,j} = b \cdot n_{gB,\,i} \cdot n_{pA,\,i} \cdot (n_{pA,\,i}-1) / 2,
$$ where $b$ is the intrinsic reaction rate, $n_{pA,\,i}$ is the copy
number of protein A in state $i$, and $n_{gB,\,i}$ is the copy number
of unbound gene B, which takes the value of 0 or 1.  For this second
order reaction, the number of possible ways of choosing two protein
molecules from $n_{pA,\,i}$ copies is $\binom{n_{pA,\,i}}{2} =
n_{pA,\,i} \cdot (n_{pA,\,i}-1) / 2$.  Transition coefficients for the
other reactions in this network can be computed similarly following this
reaction and the reactions described earlier for the represser gene
network.

{\bf MAPK network.} We consider  the second order binding reaction
$$M + MKP3
\xrightarrow{b_{14}} M\_MKP3\_Y.$$
If the ``before'' state $i$ is
transformed to the ``after'' state $j$ by one step of this reaction,
the corresponding transition coefficient $a_{i,j}$ can be computed as
$$
a_{i,j} = b_{14} \cdot n_{M,\,i} \cdot n_{MKP3,\,i}
$$ where $b_{14}$ is the intrinsic reaction rate, $n_{M,\,i}$ and
$n_{MKP3,\,i}$ are the copy numbers of M and MKP3 molecules in state
$i$, respectively. The other transition coefficients in this network
can be computed similarly using the intrinsic reaction rates given in
Fig~\ref{fig:model} and the copy numbers of reactants determined by
the ``before'' state $i$.

\section*{Authors' contributions}
JL designed the algorithm for state space enumeration and steady state
computation. YC and JL designed the method for transition matrix
generation, YC implemented these algorithms, and developed the
molecular models.  YC carried out computation. YC and JL analyzed the
data and wrote the paper together.

\newpage

\section*{Appendix}

\textbf{Algorithm 1} {\sf State Enumerator}($\bM, \bR, B$)
\begin{algorithmic}
\STATE Network model: $\bN \leftarrow \{\bM, \bR\}$;
\STATE Initial condition:  $\bS^{t=0} \leftarrow \{c_1^{0}, c_2^{0}, \ldots, c_m^{0}\}$;  Set the value of buffer capacity: $c_{m+1}^0\leftarrow B$;
\STATE Initialize the state space and the set of transitions:
    $\mathcal{X} \leftarrow \emptyset$;
    $\bT \leftarrow \emptyset$;
\STATE Stack $ST \leftarrow \emptyset; \quad$ {\tt Push}($ST,\, \bS^{t=0}$); \,
    $StateGenerated \leftarrow ${\tt FALSE}
\WHILE{$ST \ne \emptyset$}
\STATE $\bS_i \leftarrow$ {\tt Pop} $(ST)$;
\FOR{$k=1$ to $n$}
\IF{reaction $R_k$ occurs under condition $\bS_i$}
    \IF{reaction $R_k$ is a synthetic reaction and generates $u_k$ new molecules}
       \STATE $c_{m+1} \leftarrow c_{m+1} - u_k$
       \IF{$c_{m+1} \ge 0$}
          \STATE Generate state $\bS(i,\, R_k)$ that is reached by following
            reaction $R_k$ from $\bS_i$;
          \STATE $StateGenerated \leftarrow ${\tt TRUE}
       \ENDIF
    \ELSE
       \IF{reaction $R_k$ is a degradation reaction and breaks down $u_k$  molecules}
         \STATE $c_{m+1} \leftarrow c_{m+1} + u_k$
       \ENDIF
       \STATE Generate state $\bS(i,\, R_k)$ that is reached by following
            reaction $R_k$ from $\bS_i$;
       \STATE $StateGenerated \leftarrow ${\tt TRUE}
    \ENDIF
    \IF{($StateGenerated =$ {\tt TRUE}) and ($\bS(i,\, R_k) \notin \mathcal{X}$)}
       \STATE $\mathcal{X} \leftarrow \mathcal{X} \cup \bS(i,\, R_k)$;
       \STATE {\tt Push}($ST,\, \bS(i,\, R_k)$);
       \STATE $\bT \leftarrow \bT \cup t_{\bS_i,\,\bS(i,\, R_k)}$;
       \STATE  $a_{i,\,j} \leftarrow$ {\sf Transition Coefficient}($\bS_i, \, \bS(i, \, R_k), \, R_k$)
    \ENDIF
\ENDIF
\ENDFOR
\ENDWHILE
\STATE Output $\mathcal{X},\,\bT$ and $\bA = \{a_{i,\, j}\}$.
\end{algorithmic}

\textbf{Algorithm 2 }{\sf Transition Coefficient($\bS_i, \, \bS_j, \, R_k$)}
\begin{algorithmic}
\STATE Read in reaction rate parameters for $R_k$
\STATE Retrieve the copy numbers of molecular species occurring in the reaction formula of $R_k$ from the state vector $\bS_i$
\STATE  Compute the combination copy numbers of each reactant molecular species
\STATE Compute transition coefficient $a_{i,\,j}$ based on the reaction rate parameters for $R_k$, and the combination copy numbers.
\STATE Output $a_{i,j}$.
\end{algorithmic}

\newpage

\section*{Acknowledgment}
We thank Drs.\ Bhaskar DasGupta, Ming Lin, Michael Samoilov, and
Hsiao-Mei Lu for helpful discussions.  This work is supported by a
phase II 985 Project (Sub-Project:T226208001) at Shanghai Jiao Tong
University, Shanghai, China.

\bibliographystyle{unsrt}
\bibliography{net}%,bioshape,liang,lattice,chiral-flex,pot,evo}

\newpage

\begin{figure}[!ht]
\centering
\includegraphics[width=8cm]{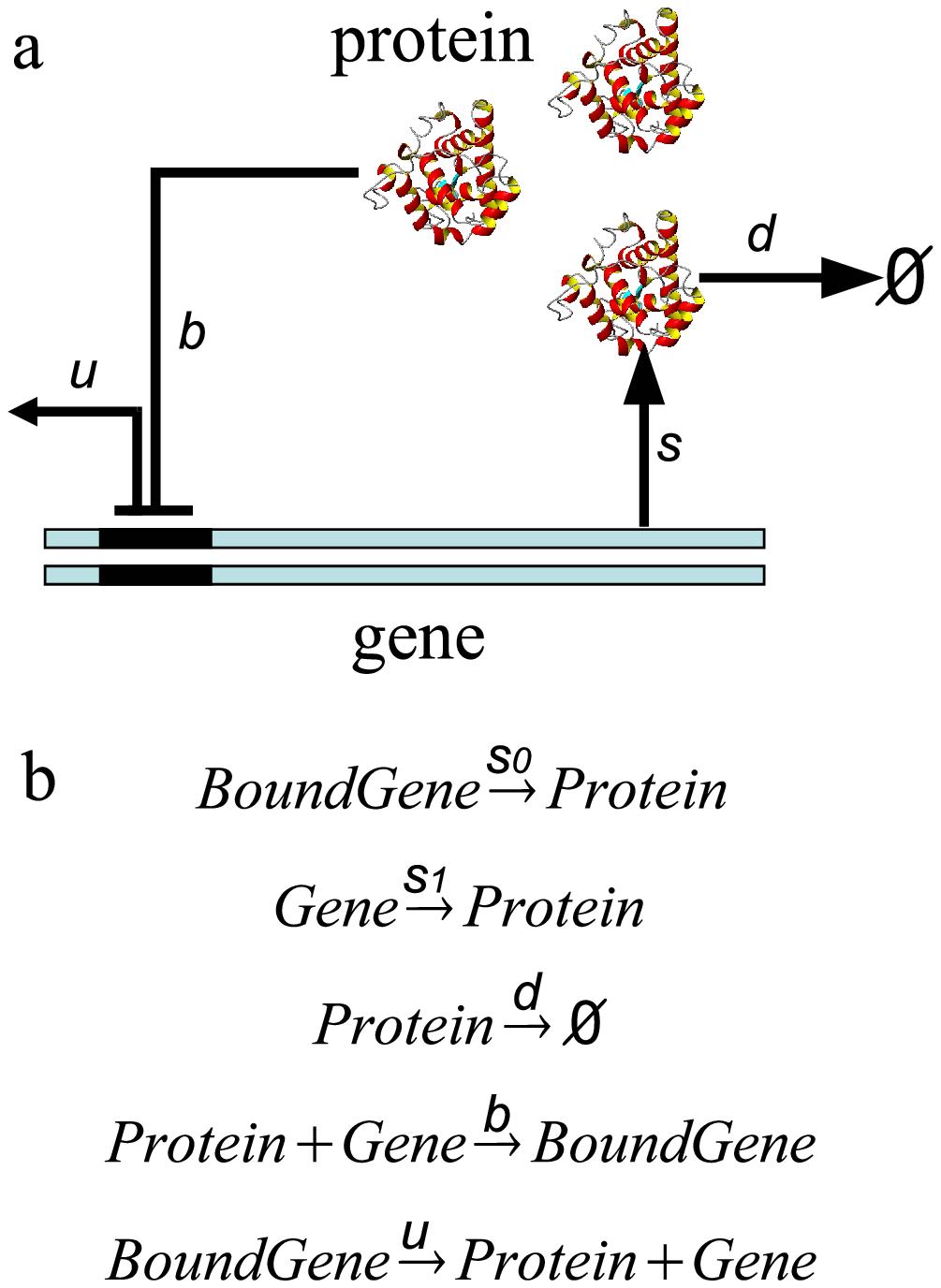}
\caption{\sf The network of a self-regulating gene. (a) The topology
of the network.  A single copy of the gene in the chromosome encodes a
protein transcription factor (TF), which is synthesized at the rate of
$s_0$ or $s_1$, depending on whether the operator site is bound (state
$0$) or unbound (state $1$).  The TF binds the operator site of the
gene at a rate of $b$.  It unbinds at a rate of $u$.  The TF is also
subject to degradation at a rate of $d$ determined by the degradation
machinery. Here the symbol $\emptyset$ represent the state of being
degraded.  (b) The chemical reactions of the five stochastic processes
and the corresponding reaction rates.}
\label{fig:selfreg-top}
\end{figure}

\begin{figure}[!ht]
\centering
\includegraphics[width=8cm]{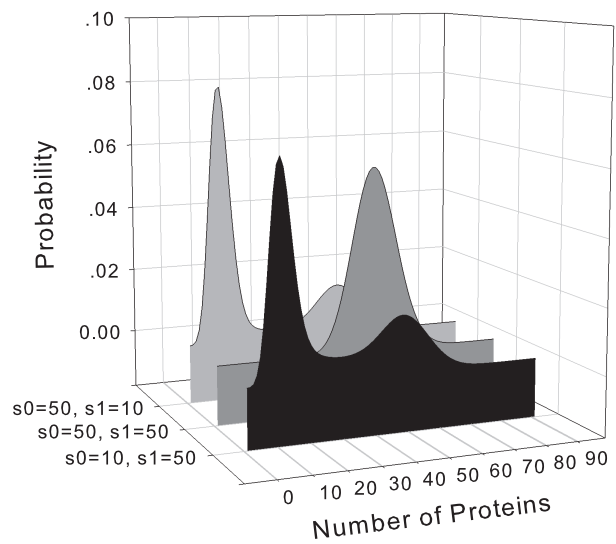}
\caption{\sf The steady state landscape probability distributions of
a self-regulating gene network. The probability over the number of free
protein is plotted.  Here this probability is the sum of probabilities
for two different gene binding states (bound and unbound) at the same
number of free proteins.  When the unbound/on state synthesis rate
$s_1$ is greater, the network is self-repressing.  When the bound/off
synthesis rate $s_0$ is greater, the network is self-activating.
Although the self-repressing (front profile) and the self-activating
(back profile) genes have overall similar distributions, the former
has a slightly higher probability in producing more free proteins than
the latter.  When both synthesis rates are equal (middle profile), the
network follows a simple birth/death process, with a Gaussian probability
distribution.  }
\label{fig:selfreg-prob}
\end{figure}

\begin{figure}[!ht]
\centering
\includegraphics[width=6.5cm]{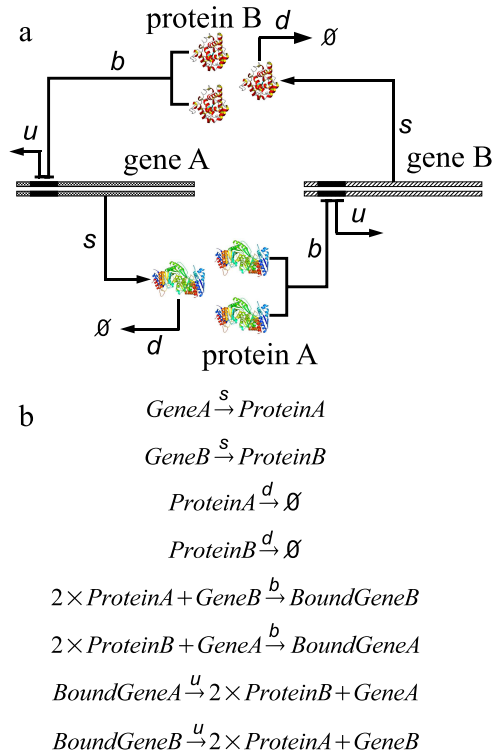}
\caption{\sf The network of a toggle switch. (a) The topology of the
network and variables representing the reaction rates.  Single copies
of gene A and gene B in the chromosome each encode a protein product.
Two protein monomers can repress the transcription of the other gene.
The synthesis of protein product of gene A and B depends on the bound
or unbound state of the gene. (b) The chemical reactions of the 8
stochastic processes involved in the toggle-switch network. The
reaction rates include $s$ for protein synthesis, $d$ for protein
degradation, $b$ for protein-gene binding, and $u$ for protein-gene
unbinding.}
\label{fig:toggle-top}
\end{figure}

\begin{figure}[!ht]
\centering
\includegraphics[width=8cm]{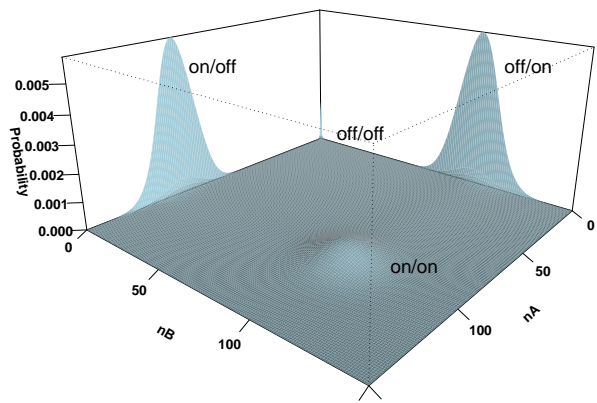}
\caption{\sf
The steady state probability landscape of a toggle switch.
A toggle switch has four different states, corresponding
to different binding state of genes A and B. At the condition of small
value of $u/b$, the off/off state is strongly suppressed,
and the system exhibits bi-stability.  }
\label{fig:toggle-prob}
\end{figure}

\begin{figure}[!ht]
\centering
\includegraphics[width=7.cm]{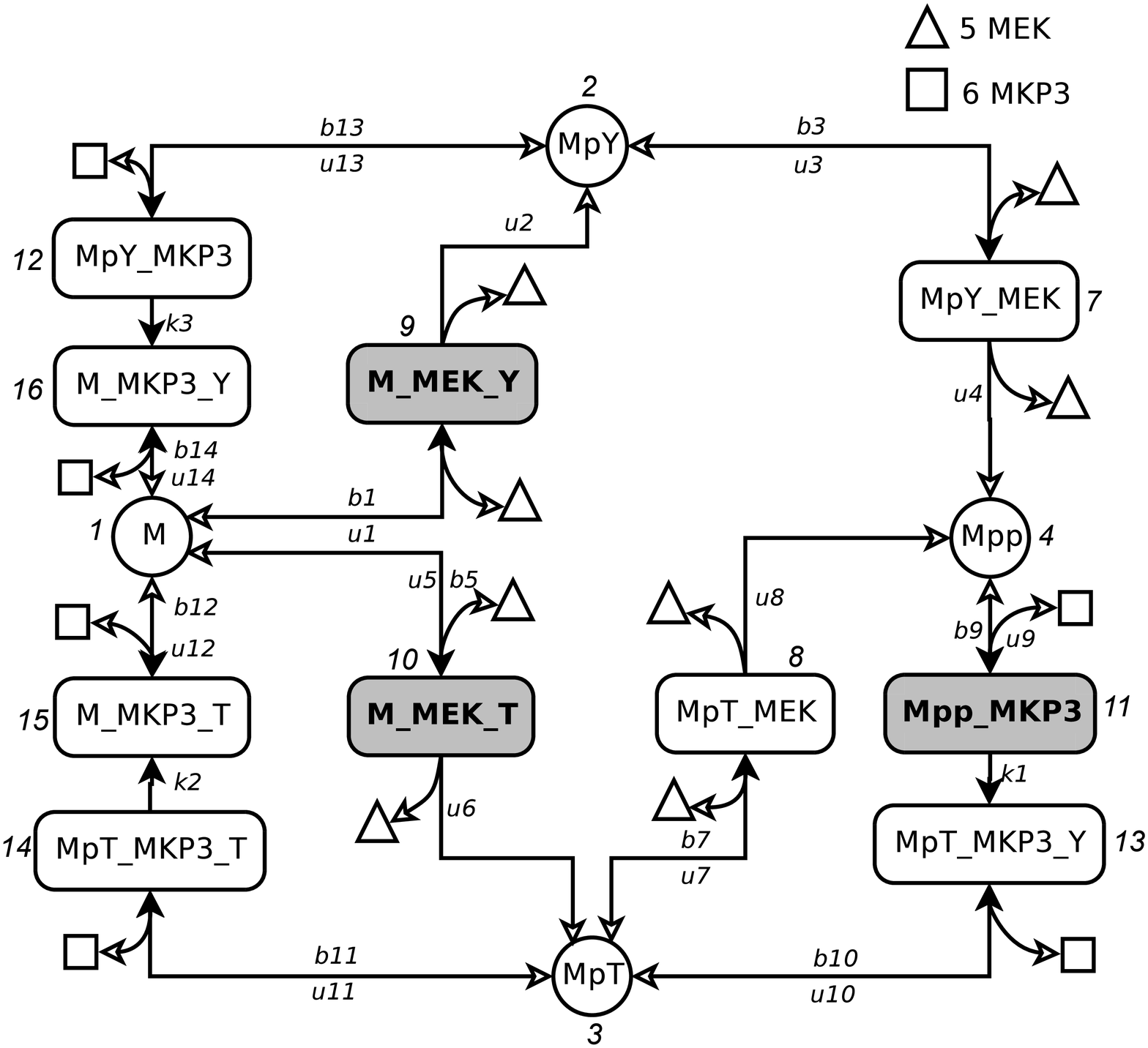}
\caption{\sf The MAPK network model according to BioModel (id
 BIOMD28).  The molecular species are labeled with integer
 numbers. Reactions are labeled with variables representing the
 corresponding reaction rate, $b_i$ for binding rates, $u_i$ for
 unbinding rates, and $k_i$ for rates of first order reactions.  Solid
 arrows in this figure represent binding reactions, and empty arrows
 for unbinding reactions.  The parameter values of this model are taken
 as is from the SBML model.  We have: $b_{1}=0.005, b_{3}=0.025,
 b_{5}=0.05, b_{7}=0.005, b_{9}=0.045, b_{10}=0.01, b_{11}=0.01,
 b_{12}=0.0011, b_{13}=0.01, b_{14}=0.0018, u_{1,3,5,7,9,10,11,13}=1,
 u_{2}=1.08, u_{4}=0.007, u_{6}=0.008, u_{8}=0.45, u_{12}=0.086,
 u_{14}=0.14, k_{1}=0.092, k_{2}=0.5$, and $k_{3}=0.47$.  }
\label{fig:model}
\end{figure}

\begin{figure}[!ht]
\centering
\includegraphics[width=8cm]{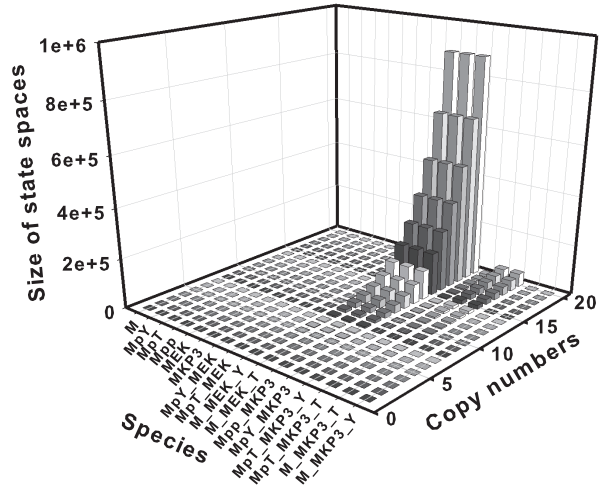}
\caption{\sf Sizes of state spaces for a model of the MAPK cascades
under the initial condition of $1$ to $20$ copies of each of the $16$
species in turn and $0$ in all other species. Altogether the size of state
space for $16\times 20 =320$ initial conditions are shown here.}
\label{fig:space}
\end{figure}

\begin{figure}[!ht]
\centering
\includegraphics[width=6.5cm]{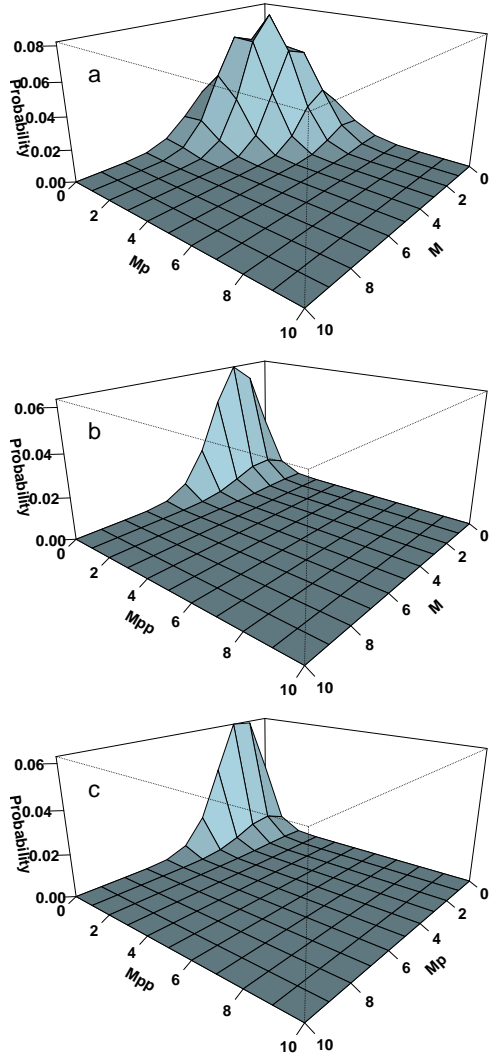}
\caption{\sf The marginal landscape probability distribution of
different copy numbers of molecular species in the MAPK network in
steady state. (a) Marginal probability distribution of the combination
of the number of unphosphorylated ERK (M) and uniphosphorylated ERK
(Mp, including both MpY and MpT), regardless of the copy numbers of
all other molecular species; (b) Marginal probability distribution of
the combination of the copy numbers of unphosphorylated ERK (M) and
dual-phosphorylated ERK (Mpp); (c) Marginal probability distribution
of the combination of the copy numbers of uniphosphorylated Mp and
dual phosphorylated Mpp.}
\label{fig:mapk-steady}
\end{figure}

\newpage

\begin{table}
\caption{\sf Abbreviations of the molecular species in the MAPK network.}
\label{tab:abbrev}
\begin{center}
\begin{small}
\begin{tabular}{|c|r|p{1.7in}|}
  \hline
  Num. & Abbrev. & Description \\
  \hline
  1 & M & ERK, extracellular signal-regulated kinase \\
  2 & MpY & ERK with Y phosphorylated \\
  3 & MpT & ERK with T phosphorylated \\
  4 & Mpp & ERK with dual phosphorylated \\
  5 & MEK & ERK kinase \\
  6 & MKP3 & ERK phosphatase \\
  7 & MpY\_MEK & Binding of MpY and MEK \\
  8 & MpT\_MEK & Binding of MpT and MEK \\
  9 & {\bf M\_MEK\_Y} & {\bf Binding of M and MEK at Y site} \\
  10 & {\bf M\_MEK\_T} & {\bf Binding of M and MEK at T site} \\
  11 & {\bf Mpp\_MKP3} & {\bf Binding of Mpp and MKP3} \\
  12 & MpY\_MKP3 &  Binding of MpY and MKP3 \\
  13 & MpT\_MKP3\_Y & Binding of MpT and MKP3 at Y \\
  14 & MpT\_MKP3\_T & Binding of MpT and MKP3 at T \\
  15 & M\_MKP3\_T & Binding of M and MKP3 at T site \\
  16 & M\_MKP3\_Y & Binding of M and MKP3 at Y site \\
  \hline
\end{tabular}
\end{small}
\end{center}
\end{table}

\end{document}